\documentclass[sigconf]{acmart}

\usepackage{algorithm}
\usepackage{algorithmic}
\usepackage{graphicx}
\usepackage{textcomp}
\usepackage{xcolor}
\usepackage{enumitem}
\usepackage{multirow}
\usepackage{booktabs}
\usepackage{diagbox}
\usepackage{colortbl}
\usepackage{subcaption}
\usepackage{verbatim}
\usepackage{textcomp} 

\AtBeginDocument{%
  \providecommand\BibTeX{{%
    \normalfont B\kern-0.5em{\scshape i\kern-0.25em b}\kern-0.8em\TeX}}}

\setcopyright{acmlicensed}

\copyrightyear{2025}
\acmYear{2025}
\setcopyright{cc}
\setcctype{by}
\acmConference[RecSys '25]{Proceedings of the Nineteenth ACM Conference on Recommender Systems}{September 22--26, 2025}{Prague, Czech Republic}
\acmBooktitle{Proceedings of the Nineteenth ACM Conference on Recommender Systems (RecSys '25), September 22--26, 2025, Prague, Czech Republic}\acmDOI{10.1145/3705328.3748045}
\acmISBN{979-8-4007-1364-4/2025/09}

\begin{document}

\title{Modeling Long-term User Behaviors with Diffusion-driven Multi-interest Network for CTR Prediction}

\author{Weijiang Lai}
\affiliation{%
\institution{Institute of Software, Chinese
Academy of Sciences}
\institution{University of Chinese Academy of Sciences
\city{Beijing}
  \country{China}}}
\email{laiweijiang22@otcaix.iscas.ac.cn}

\author{Beihong Jin$^\dagger$}
\affiliation{%
  \institution{Institute of Software, Chinese
Academy of Sciences}
\institution{University of Chinese Academy of Sciences
\city{Beijing}
  \country{China}}}
\email{Beihong@iscas.ac.cn	}

\author{Yapeng Zhang}
\affiliation{%
  \institution{Meituan}
  \city{Beijing}
  \country{China}}
\email{zhangyapeng05@meituan.com}

\author{Yiyuan Zheng}
\affiliation{%
  \institution{Institute of Software, Chinese
Academy of Sciences}
\institution{University of Chinese Academy of Sciences
\city{Beijing}
  \country{China}}}
\email{zhengyiyuan22@otcaix.iscas.ac.cn}

\author{Rui Zhao}
\affiliation{%
  \institution{Institute of Software, Chinese
Academy of Sciences}
\institution{University of Chinese Academy of Sciences
\city{Beijing}
  \country{China}}}
\email{zhaorui22@otcaix.iscas.ac.cn}

\author{Jian Dong}
\affiliation{%
  \institution{Meituan}
  \city{Beijing}
  \country{China}}
\email{dongjian03@meituan.com}

\author{Jun Lei}
\affiliation{%
  \institution{Meituan}
  \city{Beijing}
  \country{China}}
\email{leijun@meituan.com}

\author{Xingxing Wang}
\affiliation{%
  \institution{Meituan}
  \city{Beijing}
  \country{China}}
\email{wangxingxing04@meituan.com}

\thanks{$^\dagger$ Corresponding author.}


\renewcommand{\shortauthors}{W. Lai et al.}

\begin{abstract}
CTR (Click-Through Rate) prediction, crucial for recommender systems and online advertising, etc., has been confirmed to benefit from modeling long-term user behaviors. Nonetheless, the vast number of behaviors and complexity of noise interference pose challenges to prediction efficiency and effectiveness. Recent solutions have evolved from single-stage models to two-stage models. However, current two-stage models often filter out significant information, resulting in an inability to capture diverse user interests and build the complete latent space of user interests. Inspired by multi-interest and generative modeling, we propose DiffuMIN (Diffusion-driven Multi-Interest Network) to model long-term user behaviors and thoroughly explore the user interest space. Specifically, we propose a target-oriented multi-interest extraction method that begins by orthogonally decomposing the target to obtain interest channels. This is followed by modeling the relationships between interest channels and user behaviors to disentangle and extract multiple user interests. We then adopt a diffusion module guided by contextual interests and interest channels, which anchor users’ personalized and target-oriented interest types, enabling the generation of augmented interests that align with the latent spaces of user interests, thereby further exploring restricted interest space. Finally, we leverage contrastive learning to ensure that the generated augmented interests align with users' genuine preferences. Extensive offline experiments are conducted on two public datasets and one industrial dataset, yielding results that demonstrate the superiority of DiffuMIN. Moreover, DiffuMIN increased CTR by 1.52\% and CPM by 1.10\% in online A/B testing. Our source code is available at https://github.com/laiweijiang/DiffuMIN. 
\end{abstract}

\ccsdesc[500]{Information systems~Recommender systems}
\ccsdesc[500]{Information systems~Online advertising}
\ccsdesc[500]{Information systems~Learning to rank}



\keywords{CTR Prediction, User Behavior Modeling, Diffusion Model}


\maketitle

\section{Introduction}
CTR (Click-Through Rate) prediction, which estimates the probability of a user clicking a target item, has been a hot topic for both academia and industry. Recent studies and practices show that modeling long-term user behavior can improve CTR performance. 

Modeling long-term behaviors faces difficulties in effectiveness and efficiency. First, it suffers from coupled interests and noise interference, hindering accurate interest extraction. Second, the lengthy behavior sequences make it infeasible to employ mechanisms with high time complexity, such as self-attention~\cite{transformer}.


To address these challenges, current solutions have evolved from single-stage models to two-stage models~\cite{sdim,linrec,sim,eta}. Early single-stage models focus on optimizing the time complexity of target-attention or self-attention mechanisms for comprehensive modeling of long-term user behaviors. However, these simplified attention mechanisms often fail to capture user interests accurately and comprehensively. In contrast, two-stage models first filter behaviors and obtain ones that are more relevant to the target using similarity scores and then model these behaviors separately. These solutions alleviate both efficiency and noise interference issues, making them prevalent. However, despite recent advancements in two-stage approaches~\cite{ubr4ctr,cofars,twin,twin_v2}, such as incorporating richer features to aid in behavior filtering, these models tend to extract behaviors from a single perspective, resulting in homogeneity and redundancy that significantly constrain the user's interest space and ultimately limit their performance.

Compared to short-term user behaviors, long-term behaviors encompass abundant user interests. Inspired by multi-interest modeling, we propose filtering user behaviors from multiple perspectives within long-term behaviors to disentangle and extract multiple user interests. Further, although we maximize the preservation of multiple user interests through multi-interest modeling, the filtering mechanism inherently restricts the user's interest space. Inspired by generative modeling, we incorporate a diffusion module to learn the distribution of users' multiple interests, capturing nuanced and robust augmented interests. This process expands the original interests within the distribution, thereby revealing new insight for exploration and understanding user interests.

To explore user interest space and unleash the potential of long-term behaviors,  we propose DiffuMIN (Diffusion-driven Multi-Interest Network), a two-stage model for long-term behavior modeling. In the first stage, we propose a target-oriented multi-interest extraction method that begins by decomposing the target embedding into orthogonal interest channels and modeling the relationship between user behaviors and these channels. It extracts aggregated interests by routing each behavior to the channel with the highest score in relevance and aggregating behaviors with scores in the top-$p$\% in each channel, effectively reducing inter-channel redundancy and filtering out irrelevant behaviors in long-term sequences. In the second stage, we design a diffusion module to generate multiple augmented interests to supplement aggregated interests. To the best of our knowledge, we are the first to apply diffusion modeling to user interests in CTR prediction. This diffusion process is guided by contextual interests and interest channels, which anchor users' personalized and target-oriented interest types, respectively. Additionally, instead of conventional Gaussian noise, we employ perturbed user interests as the starting points for initial generation, thereby enhancing personalization and simplifying the sampling process for high-quality representations. Lastly, to further optimize generation quality, we introduce contrastive learning to ensure the augmented interests align with a user's actual preferences, enhancing the distinguishability of interests among different users.

Our contributions are summarized as follows.
\begin{itemize}
\item We propose the target-oriented multi-interest extraction method, which disentangles and extracts multiple user interests by modeling the relationship between user behaviors and interest channels derived from the orthogonal decomposition of the target, thereby deriving the diverse user interests.
\item We propose the diffusion module guided by contextual interests and interest channels, enabling the generation of augmented interests that align with the latent spaces of user interests, thereby sustaining and enriching the latent interest space.
\item We conduct extensive offline experiments on three real-world datasets and online A/B testing. Experimental results show that DiffuMIN achieves SOTA performance. 
\end{itemize}


\section{Related Work}
Our research primarily involves CTR models for long-term behaviors, multi-interest modeling, and generative modeling.

\subsection{CTR Models for Long-term Behaviors}
CTR models predict the likelihood of interactions between a user and a target. One prominent category focuses on modeling feature interactions, such as Wide\&Deep and DeepFM~\cite{cox1958regression, he2014practical, rendle2010factorization,wide, deepfm, afm, onn, dil}. Additionally, another category emphasizes analyzing user behaviors to boost CTR predictions\cite{dsin,can,bst}. The models that fall into this category include DIN \cite{din}, which utilizes an attention mechanism to prioritize relevant behaviors, DIEN \cite{dien}, which captures evolving interests with attention and GRU, and HSTU~\cite{hstu}, which integrates recall and ranking tasks within a unified architecture, exploring scaling laws in recommendation scenarios. 

Aside from these models, numerous studies focus on modeling long-term user behaviors to uncover behavior dependencies and periodic patterns within user behaviors. Early models like MIMN~\cite{mimn} and HPMN~\cite{hpmn} utilize memory networks to manage user interests but encounter challenges in providing timely updates and modeling the relationships between behaviors and targets, which ultimately limits performance. 

Recent models such as SDIM\cite{sdim} and LinRec~\cite{linrec} attempt to simplify attention mechanisms for fully long-term behavior modeling. However, these simplified approaches often struggle to effectively model the nuances of long-term behavior relationships and capture complex user behavior patterns. In contrast, other models, including SIM~\cite{sim}, begin by identifying the top-$k$ behaviors most relevant to the target item, then modeling these behaviors, respectively. Building on this two-stage line of thought, models such as UBR4CTR~\cite{ubr4ctr}, TWIN~\cite{twin}, and CoFARS~\cite{cofars} incorporate auxiliary information to improve the accuracy of filtering in the first stage \cite{ubr4ctr, twin}. Meanwhile, ETA~\cite{eta} and TWIN-v2~\cite{twin_v2} employ proven techniques such as SimHash~\cite{simhash} encoding and clustering to boost model efficiency.

Although long-term behaviors offer a wealth of user interests, current methods, particularly two-stage models, despite achieving certain results, inevitably constrain the expression of user interests, thereby limiting overall performance.

\subsection{Multi-interest Modeling Methods}
Multi-interest modeling, which enhances the performance by identifying users' diverse and dynamically changing interests, has become a focal area of research~\cite{interest1,interest2,interest3,interest4,interest5}. 

For instance, MIND~\cite{mind} and ComiRec~\cite{comirec} utilize dynamic routing and capsule networks to adaptively aggregate user behaviors into multiple embeddings, representing multiple user interests. DMIN~\cite{dmin} employs multi-head self-attention to encode user behaviors, modeling the output of each head with the target as distinct user interests. Octopus~\cite{octopus} initializes multiple orthogonal interest channels to aggregate user behaviors for multi-interest extraction. Trinity~\cite{trinity} adopts a two-clustering approach, extracting multiple user interests from the primary cluster and secondary cluster.

However, these approaches either concentrate solely on modeling relationships within user behaviors to extract multiple interests or use high-complexity techniques like self-attention, making them unsuitable for modeling long-term user behaviors in CTR scenarios.

\subsection{Generative Modeling Methods}
Common generative models include Generative Adversarial Networks (GANs), Variational Autoencoders (VAEs), and diffusion models~\cite{VAE, GAN, ddpm, stable}, with diffusion models demonstrating superior theoretical and practical performance, achieving state-of-the-art results in fields such as image generation~\cite{xia2023diffir, chung2022come}. Inspired by this, researchers try to introduce diffusion models to the recommender systems, aiming to better capture complex distributions of user behavior and features, while alleviating data sparsity issues~\cite{diffu_rec1,diffu_rec2,diffu_rec3,diffu_rec4,diffu_rec5,diffu_rec6, diffu_rec7}.

For example, DreamRec~\cite{dreamrec} employs guided diffusion to generate oracle items aligned with user interests, recommending real items that best match these oracle items. DiffRec \cite{diffrec} modifies the conventional sampling starting point from Gaussian noise to perturbed embeddings, reducing noise within the original embeddings. DiffuASR \cite{diffuasr} uses diffusion models to learn the distribution of user behavior embeddings, directly generating sequences to augment behavior data. PDRec~\cite{plug} and Diff4Rec~\cite{diff4rec} apply similar diffusion modules as DiffRec for augmentation. CaDiRec~\cite{CaDiRec} also implements diffusion to generate data guided by positional encoding and context, employing these for contrastive learning. SeeDRec~\cite{seedrec} introduces sememes as a generation granularity, enhancing existing models with additional information to boost performance. 

However, diffusion techniques have not been found to be applied to CTR prediction for modeling user interests.

\section{Methodology}

\begin{figure*}[tb]
  \includegraphics[width=\textwidth]{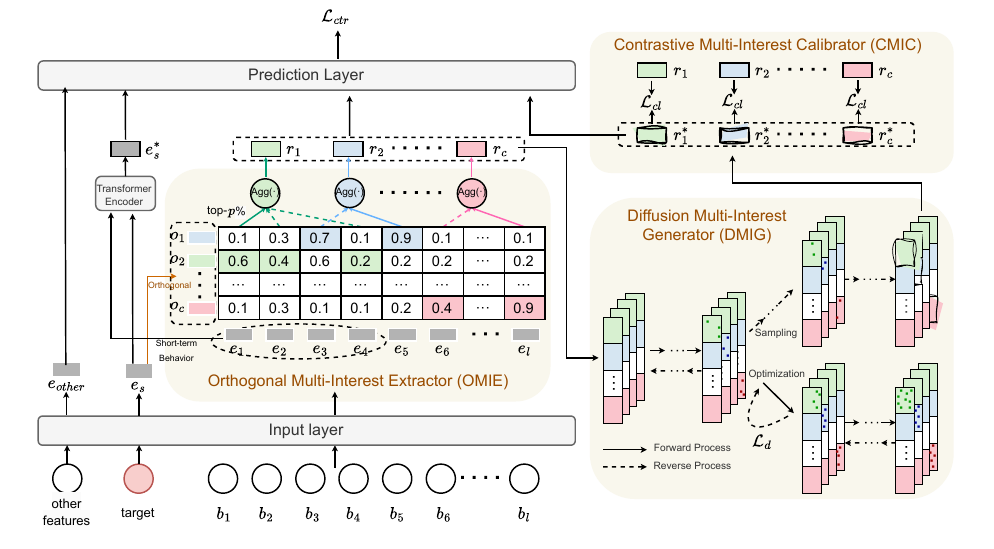}
  \caption{Architecture of DiffuMIN.}
  \label{fig:model}
\end{figure*}

\subsection{Problem Formulation}
For each user $u$, which has a user behavior sequence $\mathcal{S}=\{{b_i\}}_{i=1}^l$ and more detailed features such as age and gender. Here, $b_i$ is the $i$-th behavior and $l$ is the length of the user behavior sequence. Each behavior $b_i$ has features like item ID and category ID. Regarding a target item $s$, it includes features such as item ID and more additional contextual features.

Given a user $u$ and a target item $s$, with their interaction label $y \in \{0,1\}$, our task is to predict the click-through probability, formalized as follows:
\begin{equation}
    \mathcal{P}(y=1| x \in \{(u, s)\})=F(x; \theta),
\end{equation}
where $F(x; \theta)$ is the model we will develop and $\theta$ represents the parameters of the model.

For this task, we propose a model named DiffuMIN. Figure~\ref{fig:model} shows the architecture of our model, which mainly includes the input layer, Orthogonal Multi-Interest Extractor (OMIE), Diffusion Multi-Interest Generator (DMIG), Contrastive Multi-Interest Calibrator (CMIC), and prediction layer.

\subsection{Input Layer}
In this layer, we utilize a uniform embedding table to initialize all embeddings, including the user behavior sequence embedding matrix $E=[e_1,...,e_l]\in \mathbb{R}^{l\times d}$, where $e_i$ represents the embedding of the $i$-th behavior, the target item embedding vector $e_s\in \mathbb{R}^{d}$, and embedding vector $e_{other}$ representing other features. The embedding dimension $d$ is used uniformly to denote the dimension after the concatenation of features.

In this layer, we extract the recent $k$ behaviors to form the user short-term behavior embedding matrix $E_k = [e_1, \ldots, e_k] \in \mathbb{R}^{k \times d}$, which is input into the Transformer encoder with the target embedding $e_s$ to model the user's current interest. Simultaneously, we feed the complete behavior embedding matrix $E$ and the target embedding $e_s$ into the OMIE to disentangle and extract multiple user interests.

\subsection{Orthogonal Multi-Interest Extractor}
Compared to short-term behaviors, long-term user behaviors contain a wider range of interests. However, capturing these interests comprehensively poses a challenge due to coupled interests and numerous irrelevant behaviors within long-term behaviors. Existing two-stage models often filter behaviors based solely on their similarity to targets, leading to repetitive behaviors and interest redundancy, which substantially restrict the user's interest space.

Inspired by multi-interest modeling, we propose OMIE, designed to disentangle and extract user interests from various perspectives within long-term behaviors, mitigating the reduction of interests common in two-stage approaches. In this module, we introduce a target-oriented multi-interest extraction method utilizing the orthogonal decomposition of target embeddings as interest channels. By learning relationships between behaviors and these diverse interest channels, we decouple and extract multiple user interests, thus preserving the user's interest space. Unlike traditional multi-interest models that typically rely on neural networks to explore intra-behavior relationships for multi-interest extraction, our approach emphasizes the relationship between behaviors and the target, which is vital in CTR scenarios.

Specifically, we first perform a linear projection to the target item embedding, followed by orthogonal decomposition to derive a set of basis embeddings as interest channels. The specific formula is as follows:
\begin{equation}
e_s'=e_sW,
\end{equation}
\begin{equation}
O \leftarrow \arg\min_{O} \sum_{e_s'} \left\| e_s' - O^T O e_s' \right\|_2 ,
\end{equation}
subject to
\begin{equation}
O^T O = \mathbf{I},
\end{equation}
where $W \in \mathbb{R}^{d\times cd}$, $O = [o_1, o_2, \ldots, o_c] \in \mathbb{R}^{c \times d}$ and $c$ is the number of interest channels. Each $o_i$ represents an interest channel, capturing a specific target-oriented aspect of user interest. The orthogonality of these channels efficiently reduces interest redundancy in long-term behaviors and enhances the effectiveness of multi-interest extraction.

We determine the relationship between user behaviors and interest channels through matrix multiplication of the user behavior matrix $E$ with the interest channel matrix $O$, expressed as:
\begin{equation}
A = EO^T.
\end{equation}

In matrix $A \in \mathbb{R}^{l \times c}$, each element $A_{i,j}$ represents the score indicating the relevance between behavior $ e_i$ and channel $o_j $.

Within the long-term behavior scenario, we accurately extract multiple user interests through a process involving behavior routing, channel filtering, and interest aggregation.

\noindent\textbf{Behavior Routing.}
This step identifies the most relevant interest channel for each user behavior. We apply a top-1 routing approach, ensuring each behavior is routed exclusively to the channel where it scores the highest in relevance. This minimizes redundancy across channels, enhancing specificity and match accuracy.
\begin{equation}
    \Phi_{i,j} = 
    \begin{cases}
        1, & \text{if } j = \arg\max\limits_{j'} A_{i,j'} \\
        0, & \text{otherwise}
    \end{cases}.
\end{equation}

\noindent\textbf{Channel Filtering.}
This step aggregates only genuinely relevant behaviors within each interest channel. We filter irrelative behaviors and retain the top-$p$\% of behaviors for each channel based on their scores, thereby ensuring only the most pertinent behaviors are kept in long-term sequences, thus maintaining the accuracy of user interests.
\begin{equation}
    \Gamma_{i,j} = 
    \begin{cases}
        1, & \text{if } \Phi_{i,j}=1 \text{ and } A_{i,j} \in \text{top-}p\% \text{ of } A_{:,j} \\
        0, & \text{otherwise}
    \end{cases}.
\end{equation}

\noindent\textbf{Interest Aggregation.}
In this step, we aggregate the remaining behaviors from each interest channel to form an aggregated interest.
\begin{equation}
    r_j = \mathrm{Agg}\left( \{ e_i \mid \Gamma_{i,j}=1 \} \right),
\end{equation}
where we utilize mean pooling as the aggregation function.

\subsection{Diffusion Multi-Interest Generator}
While two-stage models effectively reduce irrelevant behaviors in long-term sequences, they concentrate on a limited subset of user behaviors to extract interests, inevitably missing out on subtle and latent user interests. Despite employing OMIE to preserve a broad range of user interests, it still inherently limits the latent space of user interests.

Inspired by generative modeling, we introduce a diffusion module to capture the distribution of multiple user interests and generate augmented interest that conforms to the latent spaces of user interests, complementing the original aggregated interest.

In traditional diffusion models, a parameterized Markov chain gradually corrupts source data with controlled noise in the forward process, transforming it into Gaussian noise. The reverse process reconstructs the data step-by-step from Gaussian noise, optimizing the network in the diffusion phase and generating data in the sampling phase.

Building on these concepts, we propose an enhanced diffusion module guided by contextual interests and interest channels. This module leverages conditional information to anchor personalized interest scopes and types effectively. Rather than starting from Gaussian noise during the sampling phase, we begin with perturbed user interests. This approach enhances personalization and streamlines the sampling process, facilitating the generation of high-quality representations.

\subsubsection{Diffusion Optimization Phase}
During the diffusion optimization phase, the diffusion network learns the distribution of each user across various interest types. In the forward process, the time step $t$ is uniformly sampled from $\{1, 2, \ldots, T\}$ to introduce noise into aggregated interest $r_{i}$, where $i \in [1,c]$. A key characteristic of the forward process is its ability to directly sample the perturbed result $r_{i,t}$ of $r_{i}$ at time step $t$ given schedule $\beta_t$, as follows:
\begin{equation}
    q\left(r_{i,t} \mid r_{i,0}\right)=\mathcal{N}\left(r_{i,t} ; \sqrt{\bar{\alpha}_t} r_{i,0},(1-\bar{\alpha}) \mathbf{I}\right),
\end{equation}
\begin{equation}
r_{i,t} = \sqrt{\bar{\alpha}_t} r_{i,0}+\sqrt{1-\bar{\alpha}_t}\epsilon,
\end{equation}
where $r_{i,0}=r_i$, $\epsilon \sim \mathcal{N}(0,1)$, $\alpha_t=1-\beta_t$, $\bar{\alpha}_t=\prod_{j=1}^t\alpha_j$, $T=1000$ and $\beta_t$ follows a linear variance schedule from 0.0001 to 0.02.

In the reverse process, the reconstruction of perturbed interest $r_{i,t}$ is guided by two conditions: the contextual interest condition $g_1=[r_1,...,r_{i-1},r_{i+1},...,r_c]$ and the interest channel condition $g_2=o_i$. The reverse process is formulated as follows:
\begin{equation}
p_\theta(r_{i,t-1}|r_{i,t},g_1,g_2)=\mathcal{N}(r_{i,t-1};\mu_\theta(r_{i,t},t,g_1,g_2),\Sigma_\theta(r_{i,t},t,g_1,g_2)),
\end{equation}

In the implementation, we only optimize the mean~\cite{ddpm}, and reconstruct $r_{i,t-1}$ from $r_{i,t}$ as follows:
\begin{equation}
    \mu_\theta(r_{i,t},t,g_1,g_2)=\frac{1}{\sqrt{\alpha_t}}(r_{i,t}-\frac{\beta_t}{\sqrt{1-\bar{\alpha}_t}}\epsilon_\theta(r_{i,t},t,g_1,g_2)),
\end{equation}
\begin{equation}
    r_{i,t-1}=\frac{1}{\sqrt{\alpha_t}}(r_{i,t}-\frac{\beta_t}{\sqrt{1-\bar{\alpha}_t}}\epsilon_\theta(r_{i,t},t,g_1,g_2))+\sigma_tz,
\end{equation}
where $z \sim \mathcal{N}(0,1)$, and $\sigma_t=\sqrt{\beta_t}$ is the standard deviation. In particular, we use the Transformer encoder as the backbone for the diffusion network $\epsilon_\theta$ to predict the added noise at step $t$ when generating $r_{i,t-1}$ from $r_{i,t}$. The diffusion network primarily incorporates self-attention and cross-attention mechanisms to support the guidance of contextual interests and interest channels.


\noindent\textbf{Contextual Interests Guidance.}
This component anchors the user's personalized interests by establishing the relationship between $r_{i,t}$ and contextual interest $g_1$ through the self-attention layer. Specifically, $r_{i,t}$ is concatenated with $g_1$ to form the matrix $R_{i,t}=[r_1,\ldots,r_{i-1},r_{i,t},r_{i+1},\ldots,r_c]$, which is then processed by the self-attention layer as follows:
\begin{equation}
Q=R_{i,t}W^Q_{self},\ K=R_{i,t}W^K_{self},\ V=R_{i,t}W^V_{self},
\end{equation}
\begin{equation}
\operatorname{Attention}(Q, K, V)=\operatorname{softmax}\left(\frac{Q K^T}{\sqrt{d}}\right) V,
\end{equation}
where $W^Q_{self}, W^K_{self}, W^V_{self} \in \mathbb{R}^{d \times d}$ are learnable projection matrices.

\noindent\textbf{Interest Channels Guidance.}
This component anchors target-oriented interest types by analyzing the relationship between user interest $r_{i,t}$ and interest channel $g_2$ using cross-attention. The specific formulas are as follows:
\begin{equation}
Q'=R_{i,t}W^Q_{cross},\ K'=g_2W^K_{cross},\ V'=g_2W^V_{cross},
\end{equation}
\begin{equation}
\operatorname{Attention}(Q', K', V')=\operatorname{softmax}\left(\frac{Q' K'^T}{\sqrt{d}}\right) V',
\end{equation}
where $W^Q_{cross}, W^K_{cross}, W^V_{cross} \in \mathbb{R}^{d \times d}$ are learnable projection matrices. In this phase, $ r_{i,t} \in R_{i,t}$ is an intermediate representation, while other embeddings $g_1$ and $g_2$ are fixed. The optimization loss for the diffusion module is calculated as follows:
\begin{equation}
    \mathcal{L}_{d}=\frac{1}{c}\sum_{i=1}^c\mathbb{E}_{r_i,t,g_1,g_2,\epsilon \sim \mathcal{N}(0,1)}[||\epsilon-\epsilon_\theta(r_{i,t},t,g_1,g_2)||_2^2].
\end{equation}

The detailed steps of the diffusion optimization phase are outlined in Algorithm~\ref{alg: optimization}.
\begin{algorithm}[tb]
    \caption{Diffusion Optimization Phase}
    \label{alg: optimization}
    \renewcommand{\algorithmicrequire}{\textbf{Input:}}
        \renewcommand{\algorithmicensure}{\textbf{Output:}}
    \begin{algorithmic}
        \REQUIRE{$r_1,...,r_c$, $o_1,...,o_c$}
                \ENSURE{$\mathcal{L}_{d}$ }

        \STATE $\mathcal{L}_{d} = 0$
        \FOR{$i \gets 1 $ to $c$ }
            \STATE $r_{i,0} \sim q(r_{i})$\\
            $t \sim \text{Uniform}(\{1,...,T\})$ \\
            $\epsilon \sim N(0,1)$ \\
            $g_1 = [r_1,...,r_{i-1},r_{i+1},...,r_c],\ \ g_2 = o_i$\\
            $\mathcal{L}_{d}+=\left\|\epsilon-\epsilon_\theta\left(\sqrt{\bar{\alpha}_t} r_{i, 0}+\sqrt{1-\bar{\alpha}_t} \epsilon, t, g_1, g_2\right)\right\|^2_2$
        \ENDFOR
    \end{algorithmic}
\end{algorithm}

\subsubsection{Diffusion Sampling Phase}
Our diffusion sampling phase diverges from traditional models that sample directly from Gaussian noise. For the $i$-th interest, we perform the forward process to obtain a perturbed interest $r_{i,t}$ as the starting point for sampling. This approach preserves user personalization while effectively removing noise from $r_i$, simplifying diffusion generation. Consequently, the reverse process, which iteratively reconstructs $r_{i,t-1}^*$ over $T'(T'<<T)$ steps, is sufficient to progressively denoise, yielding the optimized interest representation as the user's augmented interest at $t=0$, i.e., $r_i^* = r_{i,0}^*$. The formula is as follows:
\begin{equation}
r_{i,t-1}^* = \frac{1}{\sqrt{\alpha_t}}\left(r_{i,t}^* - \frac{\beta_t}{\sqrt{1-\bar{\alpha}_t}} \epsilon_\theta(r_{i,t}^*, t, g_1, g_2)\right) + \sigma_t z, \ 0 < t \leq T'.
\end{equation}

During the diffusion sampling phase, we generate $c$ different perspectives of augmented interests $r^*=[r_1^*,...,r_c^*]$ for each user, thereby enriching the corresponding aggregated interests. The specifics of the diffusion sampling phase are detailed in Algorithm~\ref{alg: Sampling}.

To achieve end-to-end training and directly enhance CTR performance through multiple augmented interests, we perform both the diffusion optimization and sampling phases during model training, while only the diffusion sampling phase is done during inference.

\begin{algorithm}[tb]
    \caption{Diffusion Sampling Phase}
    \label{alg: Sampling}
    \renewcommand{\algorithmicrequire}{\textbf{Input:}}
\renewcommand{\algorithmicensure}{\textbf{Output:}}
    \begin{algorithmic}
            \REQUIRE{$r_1,...,r_c$, $o_1,...,o_c$}
        \ENSURE{$r_{1}^*,...,r_{c}^* $ }
        \FOR{$i \gets 1 $ to $c$ }
        \STATE 
        $t \sim \text{Uniform}(\{1,...,T\})$\\
        $r_{i, t}=\sqrt{\bar{\alpha}_t}r_{i,0}+\sqrt{1-\bar{\alpha}_t}\epsilon$ \\
        $r_{i, T'}^* \sim q(r_{i, t})$\\
        $g_1 = [r_1,...,r_{i-1},r_{i+1},...,r_c], \ \  g_2 = o_i$\\
            \FOR{$t \gets T'$ to 1}
                \STATE $t>1?z \sim \mathcal{N}(0,1):z=0$ \\
                $r_{i,t-1}^*=\frac{1}{\sqrt{\alpha_t}}(r_{i,t}^*-\frac{\beta_t}{\sqrt{1-\bar{\alpha}_t}}\epsilon_\theta(r_{i,t}^*,t,g_1,g_2))+\sigma_tz$
            \ENDFOR
            \STATE $r_i^* = r_{i,0}^*$
        \ENDFOR
    \end{algorithmic}
\end{algorithm}

\subsection{Contrastive Multi-Interest Calibrator}
In OMIE and DMIG, we obtain multiple aggregated interests $r=[r_1, \ldots, r_c]$ through important behavior aggregation, and augmented interests $r^*=[r_1^*, \ldots, r_c^*]$ via diffusion sampling. To further explore the relationships between these aggregated and augmented interests, we propose employing contrastive learning to examine their similarities and differences, thereby enhancing representational learning.

Contrastive learning is a self-supervised method that achieves superior representation learning by minimizing the distance between positive samples while maximizing the separation between negative samples, thereby improving the model's generalization and robustness. In this module, for a user's $i$-th aggregated interest $r_i$, the positive sample is the user's augmented interest $r_i^*$, while negative samples are the augmented interest $r_i^*$ of other users within the mini-batch. In particular, We project these embeddings to other spaces and optimize them using the following loss function to prevent interference with the primary CTR task:
\begin{equation}
    \mathcal{L}_{cl} = \frac{1}{c|\mathcal{U}|} \sum_{u \in \mathcal{U}} \sum_{i=1}^c -\log      \frac{\exp\left( \text{sim}(r_i^{(u)}, r_i^{*(u)})/ \tau \right)}    {\sum_{v \in \mathcal{U}} \exp\left( \text{sim}(r_i^{(u)}, r_i^{*(v)})/ \tau \right)},
\end{equation}
where $\mathcal{U}$ denotes the set of users in the mini-batch and $\tau$ is the temperature. This loss function not only aligns the augmented interests with the user's actual preferences but also enhances the distinguishability of interests among different users.

\subsection{Prediction Layer}
In this layer, we input the aggregated interests $r$, augmented interests $r^*$, the user's short-term interest $e_s^*$ encoded by the Transformer encoder, and $e_{other}$ into the multilayer perceptron (MLP) for CTR prediction, as described by the following formula:
\begin{equation}
    e_s^*=\text{Transformer}([e_1,...,e_k,e_s])[-1,:],
\end{equation}
\begin{equation}
    \hat{y}=\sigma(\text{MLP}([r,r^*,e_{other},e_s^*])),
\end{equation}
where $e_1,...,e_k$ represent the embeddings of the user's short-term behaviors, $e_s^*$ is the last element of Transformer encoder output, and $\sigma$ is the sigmoid function. We use the binary cross-entropy loss to optimize our model, as follows:
\begin{equation}
    \mathcal{L}_{ctr} = -\frac{1}{N} \sum_{i=1}^N\left(y_i \log \hat{y}_i+\left(1-y_i\right) \log \left(1-\hat{y}_i\right)\right),
\end{equation}
where $N$ is the number of samples, and $y_i$ is the label of the $i$-th sample. During training, the three loss functions are optimized simultaneously as follows:
\begin{equation}
    \mathcal{L}=\mathcal{L}_{ctr}+\lambda_1 \mathcal{L}_{d}+\lambda_2 \mathcal{L}_{cl},
\end{equation}
where $\lambda_1$ and $\lambda_2$ are the weighting coefficients.

\subsection{Complexity Analysis}
\subsubsection{Space Complexity.}
In DiffuMIN, the additional learnable parameters primarily arise from the projection layers in OMIE and CMIC, and the diffusion network in DMIG. These have spatial complexities of $O(cd^2 + 2d^2)$ and $L(8d^2 + 2dd_f)$, where $L$ denotes the number of diffusion network layers and $d_f$ is the dimension of the feedforward layer in the diffusion network, respectively. Consequently, the parameter increase introduced by DiffuMIN is minimal.

\subsubsection{Time Complexity.}
The time complexity of our model is comprised primarily of OMIE, DMIG, and CMIC, with respective complexities of $O(Blcd+Blc^2)$, $O(BT'L(cd^2+c^2d+cdd_f))$, and $O(bcd^2+bc^2d)$. Given that $c<<l$ and $T'$ is small in our configuration, the overall time complexity is approximately $O(blcd)$. Thus, our model's efficiency is comparable to recent models designed for long-term user behaviors.

\section{Experiments}
In this section, we conduct extensive experiments to answer the following Research Questions (RQs):
\begin{itemize}
\item \textbf{RQ1}: Does DiffuMIN outperform existing CTR models when modeling long-term user behaviors?
\item \textbf{RQ2}: What contributions do the individual modules of DiffuMIN make to its overall performance?
\item \textbf{RQ3}: How do different designs within the OMIE module affect model performance?
\item \textbf{RQ4}: What is the performance impact of key designs in the DMIG module?
\item \textbf{RQ5}: How does DiffuMIN's interest modeling differ from traditional models in representing user interest spaces?
\item \textbf{RQ6}: How does DiffuMIN perform in the live production environment?

\end{itemize}

\begin{table}[t]
\centering
\caption{Statistics of the datasets.}
\setlength\tabcolsep{4.2pt}
\begin{tabular}{ccccccc}
\hline
Datasets        & \#Users & \#Items & \#Samples \\
\hline
Ele.me  & 14,427,689   & 7,446,116   & 128,000,000         \\
Alibaba  & 1,141,729   &  461,527  & 700,000,000         \\
Industry & 84,262,000   & 16,624,521  & 1,061,530,7879   \\
\hline
\end{tabular}
\label{tab:datsets}
\end{table}

\begin{table*}[tbp]
  \caption{Performance comparison. We conduct each experiment three times and report the average results. In each row, the best and second-best results are highlighted in bold and underlined, respectively. DIN(S) is considered the base model for calculating the RelaImpr.}
   \label{tab:performance_comparison}
  \centering
\begin{tabular}{cc|cccc|cccccc|c}
    \toprule
    Dataset & Metric & DIN(S) & CAN(S) & DIN & CAN & SoftSIM & HardSIM & ETA & SDIM & TWIN & TWIN-V2 & DiffuMIN\\
    \midrule
    \multirow{2}{*}{Industry} & AUC & 0.6740 & 0.6736 & 0.6751 & 0.6748 & 0.6772 & 0.6780 & 0.6778 & 0.6779 & 0.6785 & \underline{0.6788} & \textbf{0.6841} \\
    & RelaImpr & 0.00\% &-0.23\% & 0.63\% & 0.45\% & 1.84\% & 2.30\% & 2.18\% & 2.24\% & 2.59\%  & \underline{2.76\%} & \textbf{5.80\%}\\
    \midrule
    \multirow{2}{*}{Alibaba} & AUC & 0.6125 & 0.6091 & 0.6198 & 0.6184 & 0.6212 & \underline{0.6239} & 0.6220 & 0.6206 & 0.6215 & 0.6220 & \textbf{0.6282}\\
    & RelaImpr & 0.00\% & -3.02\% & 6.49\% & 5.24\% & 7.73\% & \underline{10.13\%} & 8.44\% & 7.20\% & 8.00\% & 8.44\% & \textbf{13.96\%} \\
    \midrule
    \multirow{2}{*}{Ele.me} & AUC & 0.6363 & 0.6378 & 0.6273 & 0.6284 & 0.6399 & 0.6389 & 0.6398 & 0.6404 & \underline{0.6410} & 0.6400 & \textbf{0.6462} \\
    & RelaImpr & 0.00\% & 1.10 \% & -6.60\% & -5.80\% & 2.64\% & 1.90\% & 2.57\% & 3.01\% & \underline{3.45\%} & 2.71\% & \textbf{7.26\%}\\
    \bottomrule
\end{tabular}
\end{table*}

\subsection{Experimental Settings}
\subsubsection{Datasets}
We select two public datasets and one industrial dataset to conduct experiments. The statistics of three datasets are shown in Table~\ref{tab:datsets}.
\begin{itemize}
\item \textbf{Alibaba}\footnote{https://tianchi.aliyun.com/dataset/56}: This dataset, provided by Alibaba, is a display advertising click-through rate prediction dataset. It includes shopping behavior data from all users over a 22-day period and includes comprehensive information on users, advertisements, and user behaviors.
\item \textbf{Ele.me}\footnote{https://tianchi.aliyun.com/dataset/131047}: 
This dataset is constructed by click logs from ele.me online recommender system and contains 30-day behaviors of users. It includes features related to users, candidate items, user behaviors, and spatiotemporal features.
\item \textbf{Industry}: It is an industrial dataset, that contains over 84 million users who have been active within the last 7 days and collects their complete behavior records for the past year, where each user behavior includes features like item ID, behavior type and so on.
\end{itemize}

\subsubsection{Competitors}
Our competitors include \textbf{DIN}\cite{din}, \textbf{CAN}\cite{can}, \textbf{SIM} \cite{sim} with its two versions \textbf{SoftSIM} and \textbf{HardSIM}, along with  \textbf{ETA} \cite{eta}, \textbf{SDIM} \cite{sdim},  \textbf{TWIN} \cite{twin} and \textbf{TWIN-V2} \cite{twin_v2}.

\subsubsection{Evaluation Metrics}
We adopt the widely used Area Under Curve (AUC) as the offline evaluation metric. For online experiments, we adopt CTR (Click-Through Rate), and CPM (Cost Per Mille) as evaluation metrics. Besides, we follow \cite{relaimpr1, din} to introduce the relative improvement (Relalmpr) metric to measure relative improvement between models, defined as follows:
\begin{equation}
    \text{RelaImpr}=(\frac{\text{AUC(model)}-0.5}{\text{AUC(base  model)}-0.5}-1) \times 100\%.
\end{equation}
\subsubsection{Implementation Details}
We implement all models by TensorFlow. For model training, we use Adam as the optimizer, and each model is trained for one epoch. Each feature dimension is set to 8. All models have the same configuration for fairness.

In DiffuMIN, we meticulously tune hyperparameters including the number of channels \( c \) within \{2, 4, 8\}, diffusion sampling step \( T' \) within \{5, 10, 20, 50\}, the temperature \( \tau \) within \{0.001, 0.005, 0.01, 0.05\}, and the weights of auxiliary losses \(\lambda_1\) and \(\lambda_2\) within \{0.0001, 0.001, 0.01, 0.1\}. Our model achieves optimal performance on the industrial dataset with \( c \), \( T' \), \( \tau \), \(\lambda_1\), and \(\lambda_2\) set to 4, 20, 0.05, 0.01, and 0.001, respectively.

On the industrial,  Alibaba, and Ele.me datasets, due to differences in data duration, we set the maximum behavior sequence length  $l$ to 5000, 1500, and 50, padding shorter sequences with zeros. In DiffuMIN, each channel aggregates the top 20\% of behaviors. The DIN and CAN models use short behavior inputs with lengths of 100, 100, and 20, resulting in DIN(S) and CAN(S). Meanwhile, two-stage models for long-term behaviors retain sequences of the second stage with lengths of 100, 100, and 20, respectively.

\subsection{Overall Performance (RQ1)}
Table~\ref{tab:performance_comparison} shows the performance results of competitors and DiffuMIN across three datasets. Note that an AUC improvement of 0.001 level is considered significant in CTR prediction scenarios \cite{twin, deepcross}. We also conduct a $t$-test on AUCs with a significance level of 0.05, indicating that there is a significant difference between DiffuMIN and comparisons in terms of performance.

The results of DIN and CAN reveal that merely extending the length of user behaviors in regular CTR models does not necessarily improve performance. More behaviors may provide more information but can overshadow crucial ones, so it is more advantageous to only model recent behaviors.

Models designed for long-term behaviors show superior ability in modeling long-term user behaviors compared to regular CTR models. Two-stage models are particularly effective, as they identify and retrieve crucial behaviors in the first stage and model them separately in the second stage. These models reduce model time complexity and mitigate irrelevant behaviors in long-term behaviors, significantly boosting performance.

DiffuMIN achieves optimal performance, highlighting the importance of extensively exploring user interests. We first derive multiple aggregated interests from various perspectives, followed by generating multiple augmented interests using a diffusion module, which substantially boosts performance.

\subsection{Ablation Study (RQ2)}
We conduct in-depth ablation experiments to analyze the contribution of each module within DiffuMIN.

Variants A and B examine the OMIE module from different perspectives. Variant A evaluates OMIE by filtering behaviors solely based on the similarity between target and behavior embeddings, then aggregating user interest and feeding it to subsequent modules. Variant B removes the multiple aggregated interests provided by OMIE in the prediction layer. Variant C omits CMIC, while Variant D removes both DMIG and CMIC. Lastly, variant E eliminates the module for modeling users' short-term behaviors.

The experimental results are presented in Table~\ref{tab:ablation}. Variant A's results demonstrate that using the target-oriented multi-interest extraction method to disentangle and extract multiple user interests effectively enhances model performance. Variant B’s AUCs indicate that explicitly capturing multiple aggregated interests are essential, with augmented interests complementing the original ones. The results of variants C and D show that our diffusion module successfully generates multiple augmented interests, thoroughly exploring users' limited interests. Furthermore, the incorporation of contrastive learning effectively boosts the expressiveness of representations. Variant E’s performance highlights the importance of extracting short-term interests, consistent with findings from other models focused on long-term behaviors.

\begin{table}
\centering
\caption{Results of the ablation study.}
\begin{tabular}{c|ccc}
\hline
Models        & Industry & Alibaba &Ele.me \\
\hline
DiffuMIN         & \textbf{0.6841}  & \textbf{0.6282}  & \textbf{0.6462}         \\
Variant A & 0.6810 &0.6257 & 0.6433 \\
Variant B & 0.6769 &0.6221 & 0.6385 \\
Variant C & 0.6834 &0.6274 & 0.6450 \\
Variant D & 0.6826 &0.6260 & 0.6448 \\
Variant E & 0.6812 &0.6250 & 0.6450 \\
\hline 
\end{tabular}
\label{tab:ablation}
\end{table}

\subsection{Analysis of OMIE (RQ3)}
In this section, we conduct experiments on the industrial dataset to analyze the OMIE module by constructing various variants. We adjust the number of channels, behavior routing, channel filtering, and interest aggregation, using DiffuMIN configured with \{4, top-1, top-20\%, mean pooling\} as the baseline.

Figure \ref{fig:OMIE_strategy} shows that selecting an optimal number of channels is critical; too few channels limit interest diversity, while too many can reduce the effectiveness of interest channel capabilities. For behavior routing, the top-1 routing method minimizes redundancy within channels, thereby boosting performance. When filtering channels, the top-$p$\% strategy is superior to top-$k$ because of the varying behavior counts across channels. Regarding interest aggregation, further aggregation using target-attention does not yield additional benefits since the relationship between the target and behaviors is already established. Although self-attention can enhance performance, its significant time demands pose challenges for online deployment.

\subsection{Analysis of DMIG (RQ4)}
\subsubsection{Advantages of Diffusion Module}
Currently, GANs~\cite{GAN} and VAEs~\cite{VAE} are prevalent generative modeling methods. Compared to GANs, VAEs are more extensively applied in recommender systems~\cite{Contrastvae, VSAN}. Thus, we focus on examining the advantages of diffusion models over VAEs in this section.

Theoretically, VAEs utilize a variational posterior to approximate the true posterior. The model's effectiveness diminishes if the variational posterior is overly simplistic, while optimization becomes challenging if the posterior is too complex. Furthermore, VAEs simultaneously optimize both the conditional distribution and the variational posterior, resulting in a large search space and issues such as posterior collapse. In contrast, diffusion models first define the variational posterior through a Markov Chain and subsequently fit it using the conditional distribution, thereby focusing solely on optimizing the conditional distribution. As a result, diffusion models provide superior and more stable performance.

Experimentally, we refer to CVAE\footnote{https://github.com/mingukkang/CVAE} and ContrastVAE\footnote{https://github.com/YuWang-1024/ContrastVAE}\cite{Contrastvae} to implement the VAE module, replacing the diffusion module in DiffuMIN to create the variant DiffuMIN-VAE.  The experimental results, presented in Table~\ref{tab:different data augmentation}, indicate that DiffuMIN-VAE does not achieve superior results on the industry and Ele.me datasets, and only approximates the results of DiffuMIN on the Alibaba dataset.

\begin{figure}[tb]
    \centering
    \begin{subfigure}[b]{0.49\linewidth}
        \includegraphics[width=\linewidth]{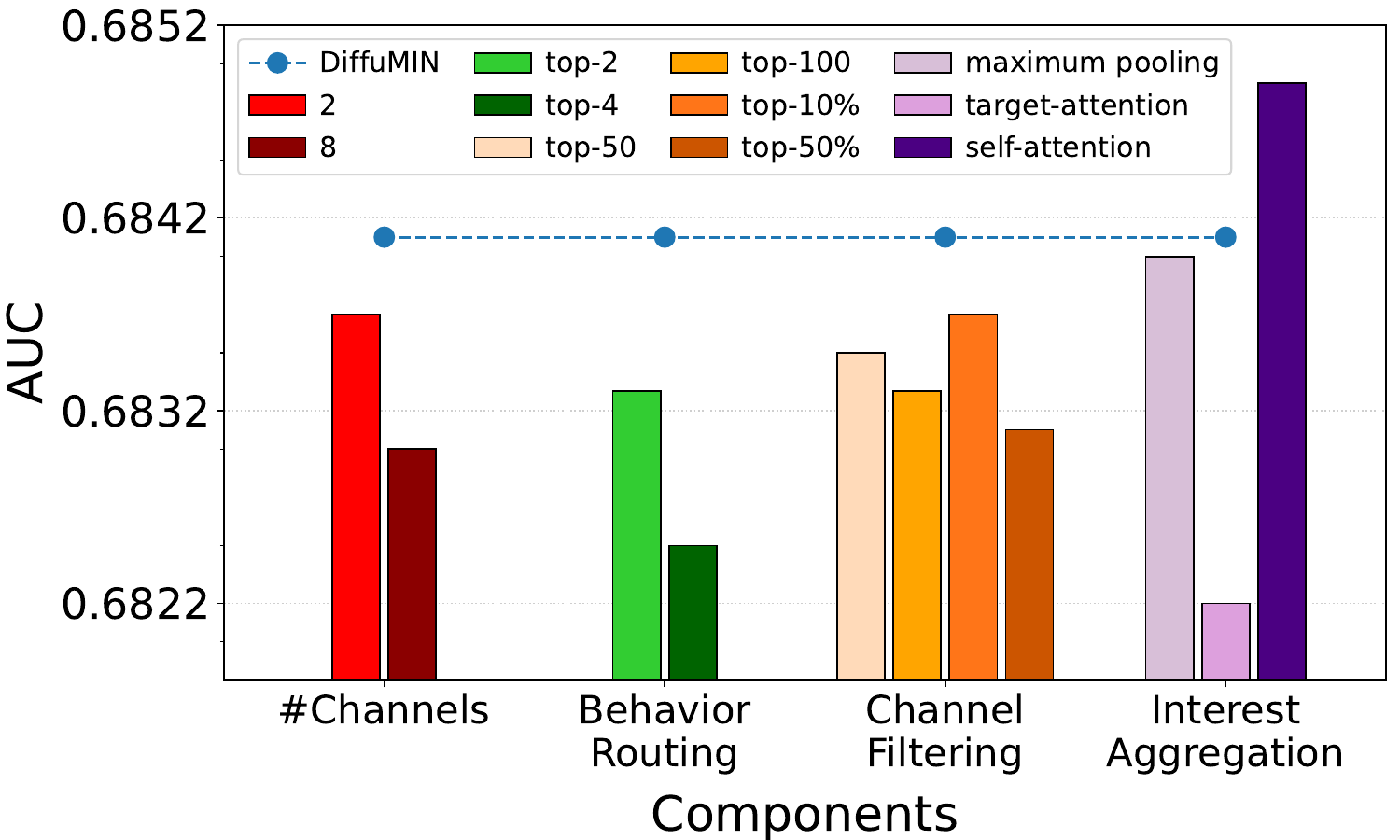}
        \caption{}
        \label{fig:OMIE_strategy}
    \end{subfigure}
    \hfill
    \begin{subfigure}[b]{0.49\linewidth}
        \includegraphics[width=\linewidth]{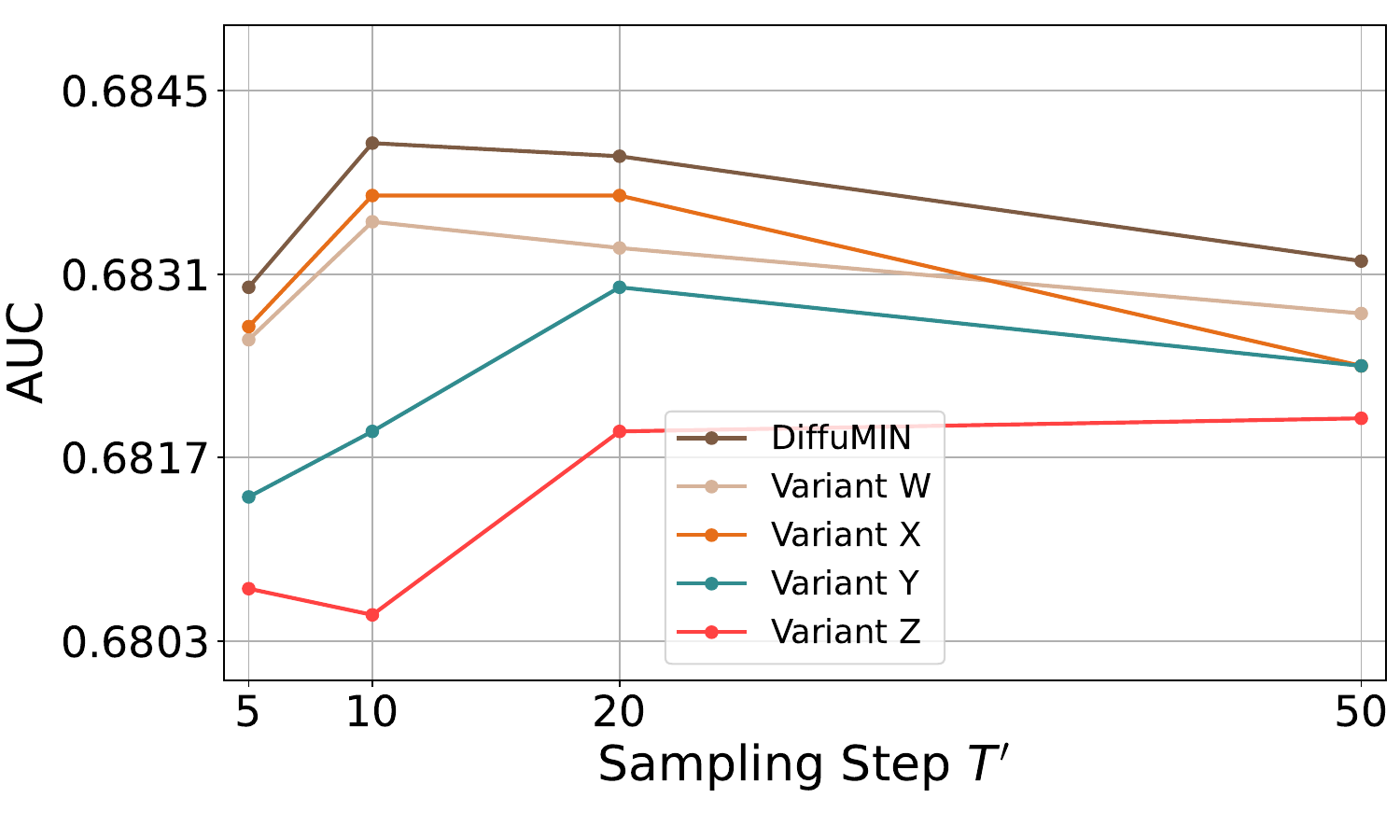}
        \caption{}
        \label{fig:comparative_performances}
    \end{subfigure}
    \caption{Comparative analysis on the industrial dataset: (a) variants of OMIE module, (b) variants of DMIG module.}
    \label{fig:combined_figures}
\end{figure}

\begin{table}
\centering
\caption{Performances with different generative modeling methods on three datasets.}
\begin{tabular}{c|ccc}
\hline
Models        & Industry & Alibaba & Ele.me  \\
\hline
DiffuMIN         & 0.6841 & 0.6282 & 0.6462     \\
DiffuMIN-VAE  & 0.6830 & 0.6277 & 0.6444           \\
\hline 
\end{tabular}
\label{tab:different data augmentation}
\end{table}
\subsubsection{Adaptation of Diffusion Module}
In our model, multiple enhancements have been applied to the traditional diffusion model. Specifically, we propose guiding the diffusion module using contextual interests and interest channels and utilizing perturbed user interests as the starting point for generation. To evaluate the effectiveness of our approach, we construct variants W, X, Y, and Z: Variant W lacks contextual interests guidance, variant X lacks interest channels guidance, variant Y omits them both, and variant Z generates directly from Gaussian noise.

The results, shown in Figure~\ref{fig:comparative_performances}, indicate that while the interests derived from OMIE offer personalized information for both user and target, removing contextual interests or interest channels guidance harms the generation quality, especially when both are absent. Variant Z's results highlight the difficulties in generating high-quality embeddings directly from Gaussian noise, necessitating additional sampling steps.

\subsection{Case Study (RQ5)}
\begin{figure}[tb]
\includegraphics[width=\linewidth]{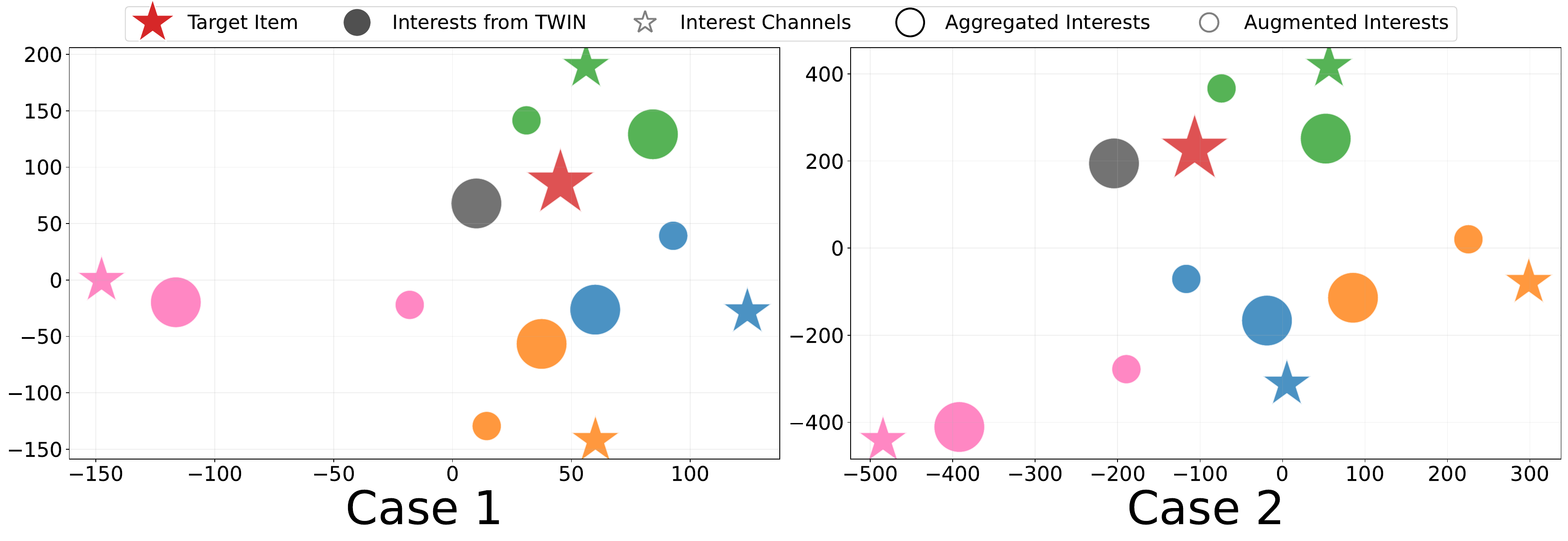}
\caption{Case study on two specific users.}
\label{fig:visual}
\end{figure}

To visually assess how DiffuMIN better preserves and explores user interest spaces compared to traditional models for long-term behaviors, we apply t-SNE to visualize the embeddings of various user interests. Figure~\ref{fig:visual} depicts two cases from the industrial dataset, each containing a user and the target clicked by the user. Large pentagrams denote targets, while small pentagrams signify interest channels decomposed from the target by using DiffuMIN. Large and small colored circles are the aggregated and augmented interests in DiffuMIN, respectively. Gray circles denote user interests derived from TWIN’s first stage. 

As illustrated in Figure~\ref{fig:visual}, each gray circle is close to a large pentagram, indicating the TWIN captures only user interest that is close to the target. However, each small pentagram is close to one or two colored circles, indicating that, in DiffuMIN, the target matches the user's interests. The reason is as follows. Our model employs the orthogonal decomposition of the target to achieve finer-grained representations as interest channels. This strategy allows us to select and aggregate user behaviors according to distinct interest channels, effectively preserving a diverse range of user interests. Furthermore, our approach extends these interests through generative methods, significantly broadening the exploration of the user's interest space.

\subsection{Online A/B Test (RQ6)}
We conducted an online A/B test deploying our DiffuMIN model in the live production environment for 7 days. The baseline model, serving the list advertisement scenario, integrates foundational models like DIN and SIM and has undergone several iterations. In the A/B test, we replaced SIM in the baseline with DiffuMIN. This led to a 1.52\% increase in CTR and a 1.10\% increase in CPM, with inference time rising slightly from 33ms to 35ms.

\section{Conclusion}
In this paper, we propose the DiffuMIN model for effectively modeling long-term user behaviors. We begin by proposing a target-oriented multi-interest extraction method to disentangle and extract multiple user interests. This is complemented by a diffusion module, guided by contextual interests and interest channels, to generate multiple augmented interests. Our approach significantly preserves and expands the constrained interest space in long-term behavior modeling, thereby enhancing the overall capacity of the model. Results from offline experiments and online A/B testing demonstrate the superiority of DiffuMIN over existing models.

\begin{acks} 
This work was supported by the National Natural Science Foundation of China under Grant No. 62072450 and Meituan.
\end{acks}

\bibliographystyle{ACM-Reference-Format}
\bibliography{sample-base}

\end{document}